# On the estimating the superconducting volume fraction from the internal magnetic susceptibility

Aleksandr V. Korolev* and Evgeny F. Talantsev**

M.N. Miheev Institute of Metal Physics, Ural Branch, Russian Academy of Sciences,
18, S. Kovalevskoy St., Ekaterinburg, 620108, Russia

* korolyov@imp.uran.ru
** evgeny.talantsev@imp.uran.ru

**Abstract**

Zhang et al.[1] reported zero-field cooled (ZFC) and field cooled (FC) data measured in a highly compressed $Pr_4Ni_3O_{10}$ single crystal. These measurements provide unambiguous confirmation of bulk superconductivity in pressurized Ruddlesden-Popper nickelates. Zhang et al.[1] applied an equation (described in Refs.[2,3]) to recalculate ZFC data measured in a $Pr_4Ni_3O_{10}$ (sample S3) in volume fraction $f$ of the superconducting phase in the sample. In result[1], $f = 0.85$ was reported for sample S3 at pressure $P = 40.2$ GPa. The key postulate of the methodology for the calculation of $f$ (see also works[4–6]) is that $f$ is equal to the amplitude of the internal magnetic susceptibility $|\chi_{internal}|$, or $f = |\chi_{internal}|$. Here we argue that this postulate is incorrect and present counterexample where the $Pr_4Ni_3O_{10}$ sample S3 can exhibit $f < 0.10$ and $|\chi_{internal}| > 0.80$. In the result, we addressed recent Replies[2,3] on our Comments[7,8]. Considering that the postulate $f = |\chi_{internal}|$ is widely used in superconductivity, we extend our request to reconsider the validity of this postulate in the entire field of superconductivity.

We acknowledge the prompt and extended Replies[2,3] of the authors[1,4–6] on our Comments[7,8], as well as our previous open-mind exchange of opinions on the topic.

In particular, in their Replies[2,3] the authors[1,4–6] stated that their calculations of the superconducting volume fraction $f$ in the sample are based on magnetic susceptibility values $\chi$ rather than on the values of the ratio of the measured magnetic moment of the sample in ZFC mode, $m_{ZFC,measured}$, to the magnetic moment of the sample in the Meissner state, $m_{Meissner}$, which we used in our Comments[7,8].

In response to this viewpoint, here we present our consideration by using the magnetic susceptibility values $\chi$.

First, we consider one report[9] from the reference lists presented in Replies[2,3]. Jiang et al.[9] (as well as authors[1–6]) used the equation (which we converted to SI units):

$$\chi_{internal} = \frac{\chi_{experimental}}{1 - N \times \chi_{experimental}} \qquad (1)$$

where $\chi_{experimental} = -\frac{|m_{ZFC,measured}|}{|H_{applied} \times V|}$, $H_{applied}$ is the applied magnetic field, $V$ is the sample volume, $N$ is the demagnetization factor of the sample[10–14], and $\chi_{internal}$ is the calculated value, that is also termed as superconducting shielding fraction[9], and which in accordance with reports[1–6]:

$$f = |\chi_{internal}| \qquad (2)$$

Jiang et al.[9] studied disk-shaped single crystal $Ca_{0.74(1)}La_{0.26(1)}(Fe_{1-x}Co_x)As_2$ (x = 0.025) that has the ratio of the disk height $c$ to the disk radius $r$, $\frac{c}{r} = 0.10$. Magnetic field in the experiment[9] was applied in a parallel direction to the disk radius. For this geometry, an approximated formula for $N$ is provided in Ref.[9]:

$$N = \frac{\pi}{2} \times \frac{c}{r} \qquad (3,a)$$

$$N = \frac{\pi}{2} \times \frac{c}{r} = \frac{3.1416}{2} \times \frac{1}{10} = 0.157 \qquad (3,b)$$

Jiang and co-workers[9] reported measured value:

$$\chi_{experimental} = -0.90, \qquad (4)$$

this can be seen in their Figure 1,d[9] for the sample with x = 0.025. The substitution of values from Equations 3,b and 4 in Equation 1 yields:

$$\chi_{internal} = \frac{\chi_{experimenta}}{1 - N \times \chi_{experimental}} = \frac{(-0.90)}{1 - 0.157 \times (-0.90)} = -0.788 \cong -0.79 \qquad (5)$$

Jiang et al.[9] reported the rounded value of:

$$\chi_{internal} = -0.80, \qquad (6)$$

or

$$f = |\chi_{internal}| = 0.80, \qquad (7)$$

which is the key postulate of published reports[1–6]. This means that the superconducting volume fraction in $Ca_{0.74(1)}La_{0.26(1)}(Fe_{1-x}Co_x)As_2$ (x = 0.025) is 80%.

Let us apply the same routine to the experimental ZFC data on a highly compressed $Pr_4Ni_3O_{10}$ single crystal (sample S3) reported by Zhang et al.[1]. Zhang et al.[1] reported that sample S3 had a disk shape with a diameter $d = 210$ μm and a thickness $h = 25$ μm. Magnetic field in experiments by Zhang et al.[1] was applied to parallel direction to the sample thickness. By substituting these parameters in the equation for the demagnetization factor $N$ for this geometry[12]:

$$N = 1 - \frac{1}{1 + q \times \frac{d}{h}}, \qquad (8)$$

$$q = \frac{4}{3\pi} + \frac{2}{3\pi} \times tanh\left(1.27 \times \frac{h}{d} \times ln\left(1 + \frac{d}{h}\right)\right), \qquad (9)$$

one can obtain for sample S3:

$$N = 0.8057. \qquad (10)$$

In Figure 1 we showed the reported[1] data for $m_{ZFC,measured}\left(P = 40.2 \ GPa, H = 1591.5 \ \frac{A}{m}\right)$ (from Figure 5,a[1]) together with:

$$m_{Meissner}\left(P = 40.2 \ GPa, H = 1591.5 \ \frac{A}{m}\right) = -6.00 \times 10^{-9} \ Am^2. \qquad (11)$$

The latter value was calculated by the routine described in Appendix A (for these calculations, we used the method described in our Comments[7,8]. We also should acknowledge that the authors[1–6] agreed with our calculations[7,8] of the $m_{Meissner}$ for all previously analyzed cases).

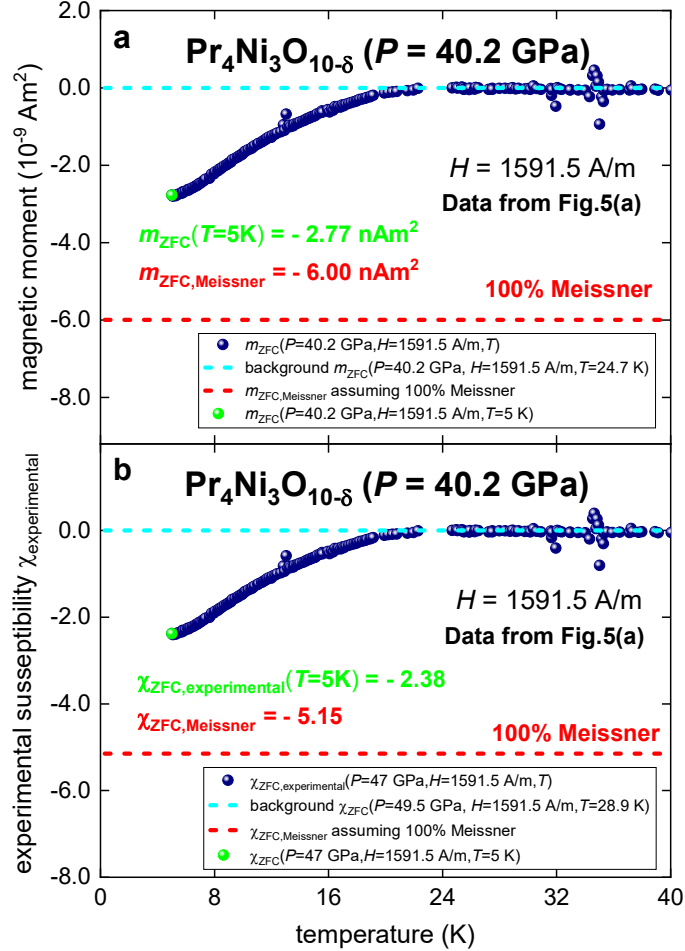

**Figure 1.** ZFC magnetization data reported by Zhang et al.[1] for $Pr_4Ni_3O_{10}$ sample S3. (a) ZFC data represented in the form of measured magnetic moment of the sample $m_{ZFC,measured}$ $\left(P = 40.2 \text{ GPa}, H = 1591.5 \frac{A}{m}, T = 5 \text{ K}\right)$. (b) ZFC data represented in the form of experimental susceptibility $\chi_{ZFC,experiemntal}$ $\left(P = 40.2 \text{ GPa}, H = 1591.5 \frac{A}{m}, T = 5 \text{ K}\right)$. The values calculated in the assumption of Meissner state of the sample are shown (in red).

One can see that at $P = 40.2$ GPa and $T = 5$ K, the measurements show that:

$$m_{ZFC,measured}\left(P = 40.2 \text{ GPa}, H = 1591.5 \tfrac{A}{m}, T = 5 \text{ K}\right) = -2.77 \times 10^{-9} \text{ Am}^2.$$

(12)

From this magnetic moment value, one can calculate:

$$\chi_{experimental}\left(P = 40.2 \text{ GPa}, H = 1591.5\ \frac{\text{A}}{\text{m}}, T = 5 \text{ K}\right) = -\frac{|m_{ZFC,measured}|}{|H_{applied} \times V(P=40.2 \text{ GPa})|} =$$

$$-\frac{|-2.77 \times 10^{-9} \text{ Am}^2|}{\left|\left(1591.5\ \frac{\text{A}}{\text{m}}\right) \times \left(\left(\frac{\pi}{4} \times (2.1 \times 10^{-4})^2 \times (2.5 \times 10^{-5})\ m^3\right) \times 0.845\right)\right|} = -2.38. \qquad (13)$$

where we used experimental data on the pressure dependence of the unit cell volume

$\frac{V_{unit\ cell}(P=40.2\ \text{GPa})}{V_{unit\ cell}(P=0\ \text{GPa})} = 0.845$ reported by Zhang et al.[1] for $Pr_4Ni_3O_{10}$ phase. Here we also accepted the assumption[1–8] that the demagnetization factor $N$ of the sample is not changed under high-pressure conditions.

The substitution of the values from Equations 10,13 in Equation 1 yields:

$$\chi_{internal}\left(P = 40.2 \text{ GPa}, H = 1591.5\ \frac{\text{A}}{\text{m}}, T = 5 \text{ K}\right) = \frac{\chi_{experimental}}{1 - N \times \chi_{experimental}} =$$

$$\frac{(-2.38)}{1 - 0.8057 \times (-2.38)} = -0.816. \qquad (14)$$

The obtained value of $\chi_{internal}$, in accordance with the postulate of the reports[1–6], means that sample S3 contains 82% of superconducting volume fraction:

$$f = |\chi_{internal}| = 0.816 \cong 0.82. \qquad (15)$$

Zhang et al.[1] reported slightly higher value $f = |\chi_{internal}| = 0.85$, and the difference originates from the accuracy of calculations of the demagnetization factor $N$ for which we used Equations 8,9, while Zhang et al.[1] used approximated equation[15].

The crucial question here is: where is 18% (or ~ 1/5 part) of non-superconducting volume fraction located in the $Pr_4Ni_3O_{10}$ sample S3? This is not an unimportant question because the authors[1–6] argued to our Comments[7,8] that their single crystals of Ruddlesden-Popper nickelates are uniform and high quality across the entire sample volume. But the fifth part of the sample is *de facto* non-superconducting in sample S3.

For $La_2SmNi_2O_7$ samples fabricated and studied by Li et al.[5] this lost part is ~ 40% (in full accordance with Equation 1). Where is almost half of the sample volume (which is non-superconducting) located, if the samples[1–6] are uniform and a high quality across the entire volume?

As we show in Ref.[7] that particular location of 50% of the non-superconducting part within the sample volume changes the sample demagnetization factor $N$, making all calculations using Equations 1,2 unfounded.

Let us show this problem again by the use of the internal susceptibility value $\chi_{internal}$. In particular, let us consider a case when 90.2% of the Ruddlesden-Popper nickelate $Pr_4Ni_3O_{10}$ sample S3[1] is not a superconductor. Thus, a Sample A with physical sizes as the $Pr_4Ni_3O_{10}$ sample S3[1]:

$$d = 210 \ \mu m \tag{16}$$

$$h = 25 \ \mu m \tag{17}$$

has the superconducting part in the form of a disk shape lamella:

$$d_{supercond} = 178.3 \ \mu m \tag{18}$$

$$h_{supercond} = 3.4 \ \mu m \tag{19}$$

$$V_{supercond} = \frac{\pi}{4} \times (1.783 \times 10^{-4})^2 \times (3.4 \times 10^{-6}) \ m^3 = (8.49 \times 10^{-14}) \ m^3 \tag{20}$$

$$f_{Sample\_A} = \frac{V_{supercond}}{V_{physical}} = \frac{\frac{\pi}{4} \times (1.783 \times 10^{-4})^2 \times (3.4 \times 10^{-6}) \ m^3}{\frac{\pi}{4} \times (2.1 \times 10^{-4})^2 \times (2.5 \times 10^{-5}) \ m^3} = 0.098 = 9.8\% \tag{21}$$

Substituting the values from Equations 18,19 in Equations 8,9, the demagnetization factor is:

$$N = 0.9589. \tag{22}$$

Magnetic moment of this sample at $P = 40.2$ GPa and applied field $H = 1591.5 \ \frac{A}{m}$ is:

$$m_{Sample\_A,measured}\left(P = 40.2 \ \text{GPa}, H = 1591.5 \ \tfrac{A}{m}\right) = -V_{supercond} \times$$

$$\frac{V_{unit \ cell}(P=40.2 \ \text{GPa})}{V_{unit \ cell}(P=0 \ \text{GPa})} \times \frac{H}{1-N} = -\left((8.49 \times 10^{-1}) \ m^3\right) \times 0.845 \times \frac{1591.5 \ \tfrac{A}{m}}{1-0.9589} = -2.77 \times 10^{-9} \ Am^2, \tag{23}$$

which is the exact value as the measured value (Equation 12 and Figure 1,a) for $Pr_4Ni_3O_{10}$ sample S3 by Zhang et al.[1].

Let us calculate $\chi_{experimental}\left(P = 40.2 \text{ GPa}, H = 1591.5 \frac{A}{m}\right)$ for Sample A:

$\chi_{experimental,Sample\_A}\left(P = 40.2 \text{ GPa}, H = 1591.5 \frac{A}{m}\right) =$

$-\frac{|m_{Sa\_A,measured}|}{|H_{applied} \times V_{physical}(P=40.2 \text{ GPa})|} = -\frac{|-2.77 \times 10^{-9} \text{ Am}^2|}{\left|\left(1591.5 \frac{A}{m}\right) \times \left(\left(\frac{\pi}{4} \times (2.1 \times 10^{-4})^2 \times (2.5 \times 10^{-5}) \text{ m}^3\right) \times 0.845\right)\right|} = -2.38$

(24)

This value is again identical to the one for $Pr_4Ni_3O_{10}$ sample S3[1]. (eq. 13, and Figure 1,b). Substituting the values from Equations 10,24 into Equation 1 results in:

$\chi_{internal,Sample\_A}\left(P = 40.2 \text{ GPa}, H = 1591.5 \frac{A}{m}\right) = \frac{\chi_{experimental,Sample\_A}}{1 - N \times \chi_{experimental,Sample\_A}} =$

$\frac{(-2.38)}{1 - 0.8057 \times (-2.38)} = -0.816 \cong -0.82.$ (25)

Following Equation 2:

$f = |\chi_{internal,Sample\_A}| = 0.82.$ (26)

Equation 26 contradicts Equation 20:

$f_{Sample\_A} = 0.098 \ll |\chi_{internal,Sample\_A}| = 0.82.$ (27)

Thus, we present our response to Replies[2,3] by demonstrating that the calculations of $\chi_{internal}$ do not solve the problem, and the superconducting volume fraction remains uncertain.

We should clarify that our critique[7,8] has a broader context, and it does not limit to pressurized Ruddlesden-Popper nickelates, but applies to all superconductors for which ZFC and FC data are analysed (extended reference lists are given in Refs.[2,3]).

We therefore need to call for the re-evaluation of the validity of the postulate (Equations 1,2) as a methodology used in superconductivity.

## Author contributions

## Acknowledgements

The work was carried out within the framework of the state assignment of the Ministry of Science and Higher Education of the Russian Federation for the IMP UB RAS.

## Competing interests

*Appendix A*

**On the estimating the superconducting volume fraction from the internal magnetic susceptibility**

Aleksandr V. Korolev[*] and Evgeny F. Talantsev[**]

M.N. Miheev Institute of Metal Physics, Ural Branch, Russian Academy of Sciences,
18, S. Kovalevskoy St., Ekaterinburg, 620108, Russia

* korolyov@imp.uran.ru
** evgeny.talantsev@imp.uran.ru

A superconductor of any physical shape in the Meissner state[16] exhibits zero magnetic flux density within its material, $B = 0$. Based on that, for a sample in the Meissner state[16] with magnetometric[17] demagnetization factor[12,15,18,19] $N$, the primary equation of the magnetostatic is:

$$B = 0 = \mu_0 \times (M_V + H - N \times M_V), \tag{A1}$$

$$M_V = -\frac{H}{1-N}, \tag{A2}$$

where $\mu_0$ is the magnetic permeability of free space (in units of $[NA^{-2}]$), $H$ is the magnetic field (in units of $\left[\frac{A}{m}\right]$), and $M_V$ is the volume magnetization of the sample (in units of $\left[\frac{A}{m}\right]$). The latter value is calculated as a ratio of the measured in the experiment magnetic moment of the sample $m$ (in units of $[Am^2]$) and the sample volume $V$ (in units of $[m^3]$):

$$M_V = \frac{m}{V}. \tag{A3}$$

Because under high-pressure conditions the sample volume changes, even if the measured magnetic moment of the sample does not change, the volume magnetization value will change. Therefore, experimental pressure dependence of the unit cell volume $V_{u.c.}(P)$ is essential experimental data for ZFC and FC data analysis.

And, if for ambient pressure samples, the combining Equations A1-A3 yields:

$$m_{Meissner} = -V_0 \times \frac{H}{1-N}, \tag{A4}$$

where $V_0$ is the sample volume at ambient pressure, the magnetic moment of the pressurized sample in the Meissner state is described by the following equation:

$$m_{Meissner}(P, H) = -V_s(P) \times \frac{H}{1-N} = -V_0 \times \frac{V_{u.c.}(P)}{V_{0,u.c.}} \times \frac{H}{1-N} \tag{A5}$$

where $V_{0,u.c.}$ is the unit cell volume at ambient conditions. For the $Pr_4Ni_3O_{10}$ phase Zhang et al.[1] reported $\frac{V_{unit\,cell}(P=40.2\,\text{GPa})}{V_{unit\,cell}(P=0\,\text{GPa})} = 0.845$.

The demagnetization factor $N = 0.8057$ for sample S3[1] is calculated by Equations 8,9 in the main text. Thus, in the applied field $H = 1591.5\,\frac{A}{m}$ and pressure $P = 40.2\,GPa$ (these conditions are applied for sample S3[1]) the Meissner magnetic moment of the sample S3[1] is:

$$m_{Meissner}\left(P = 40.2\,GPa, H = 1591.5\,\frac{A}{m}\right)[Am^2] = -V_0 \times \frac{V_{u.c.}(P)}{V_{0,u.c.}}[m^3] \times \frac{H\left[\frac{A}{m}\right]}{1-N} =$$

$$-\left(\frac{\pi}{4} \times (2.1 \times 10^{-4})^2 \times (2.5 \times 10^{-5})\,m^3\right) \times 0.8057 \times \frac{1591.5\,\frac{A}{m}}{1-0.8057} = -6.00 \times 10^{-9}\,Am^2.$$

$$\tag{A6}$$

From the Meissner magnetic moment of the sample S3[1], one can calculate experimental susceptibility for sample S3[1] in the Meissner state:

$$\chi_{experimental}\left(P = 40.2\,\text{GPa}, H = 1591.5\,\frac{A}{m}\right) = -\frac{|m_{ZFC,measured}|}{|H_{applied} \times V(P=40.2\,\text{GPa})|} =$$

$$-\frac{|-6.00 \times 10^{-9}\,Am^2|}{\left|\left(1591.5\,\frac{A}{m}\right) \times \left(\left(\frac{\pi}{4} \times (2.1 \times 10^{-4})^2 \times (2.5 \times 10^{-5})\,m^3\right) \times 0.845\right)\right|} = -5.15. \tag{A7}$$